\documentclass[
 reprint,amsmath,amssymb,
 aps]{revtex4-1}

\usepackage{footnote}
\usepackage{graphicx}
\usepackage{dcolumn}
\usepackage{bm}
\usepackage{float}

\usepackage[utf8]{inputenc}
\usepackage{subfigure}

\usepackage[spanish,es-tabla]{babel}

\renewcommand{\date}{Fecha}

\begin{document}

\title{Estudio del crecimiento del Sn/Ag(111): explorando la posibilidad de formación de “estaneno”}

\author{Lucas Daguerre}
\affiliation{Laboratorio Avanzado,  Instituto Balseiro, Bariloche, R8402AGP, Argentina\\E-mail:  \href{mailto:lucas.daguerre@ib.edu.ar}{lucas.daguerre@ib.edu.ar};\:\:\href{mailto:lucasdaguerreunlp@gmail.com}{lucasdaguerreunlp@gmail.com}}
\author{Director: Hugo Ascolani}
\affiliation{Laboratorio de Superficies, Centro Atómico Bariloche, Bariloche, R8402AGP, Argentina\\E-mail:  \href{mailto:ascolani@cab.cnea.gov.ar}{ascolani@cab.cnea.gov.ar}}
\author{Junio, 2019}

\begin{abstract}
En los últimos años los materiales 2D han captado la atención de la comunidad científica debido a sus propiedades físicas superlativas. El “estaneno”, un compuesto análogo al grafeno pero a base de átomos de Sn, podría poseer propiedades únicas debido al intenso acoplamiento spin-órbita (como el efecto Hall Cuántico de Spin QSH, superconductividad topológica, entre otras) que podrían tener eventuales aplicaciones en la spintrónica y la computación cuántica. En el experimento se exploró la posibilidad del crecimiento epitaxial de “estaneno” al evaporar Sn sobre un sustrato de Ag(111). Resultados parciales coinciden con la literatura utilizando técnicas de espectroscopía como LEED y XPS. En cuanto a las mediciones con UPS/ARPES se obtuvieron las relaciones de dispersión para el estado de superficie de la muestra de Sn/Ag(111): para 1/3MC de Sn la relación obtenida reprodujo los resultados reportados para la aleación de superficie Ag$_2$Sn; por el contrario, para (1/3+0,5)MC de Sn no se obtuvo la relación de dispersión parabólica reportada para el “estaneno”, más bien una relación que resultó indistinguible de la correspondiente a la aleación de superficie. Muy posiblemente, la falta de formación de “estaneno” se debió  a un calentamiento excesivo de la muestra durante su preparación lo cual habría causado  un aumento de espesor de la aleación Ag$_2$Sn.
\newline
\newline
$\:\:\:$ In recent years, 2D materials have attracted increasing attention from the scientific community due to their superlative properties. The “stanene”, a graphene like compound formed by Sn atoms, may have unique properties because of the spin-orbit coupling SOC (such as the Quantum Spin Hall Effect QSH, topological superconductivity, among others), that could eventually have applications in spintronics and quantum computing. In the experiment “stanene” epitaxial growth was explored by evaporating Sn onto a Ag(111) substrate. Partial results using spectroscopic techniques such as LEED and XPS agreed with literature. Regarding measurements made with UPS/ARPES, dispersion relations were obtained for the surface state of the Sn/Ag(111) sample: for 1/3ML of Sn they matched with the reported results for the surface alloy Ag$_2$Sn; conversely, for (1/3+0,5)ML of Sn they mismatched with the reported parabolic relation for the “stanene”, particularly, they were indistinguishable from the surface alloy one. Probably, lack of formation of “stanene” was caused by an excessive heating during the sample preparation process, that could have produced an increase in the surface alloy Ag$_2$Sn thickness.  
\end{abstract}

\pacs{Valid PACS appear here}
\maketitle


\section{\label{sec:intro}Introducción}

\begin{figure}[ht]
    \centering
    \includegraphics[width=\linewidth]{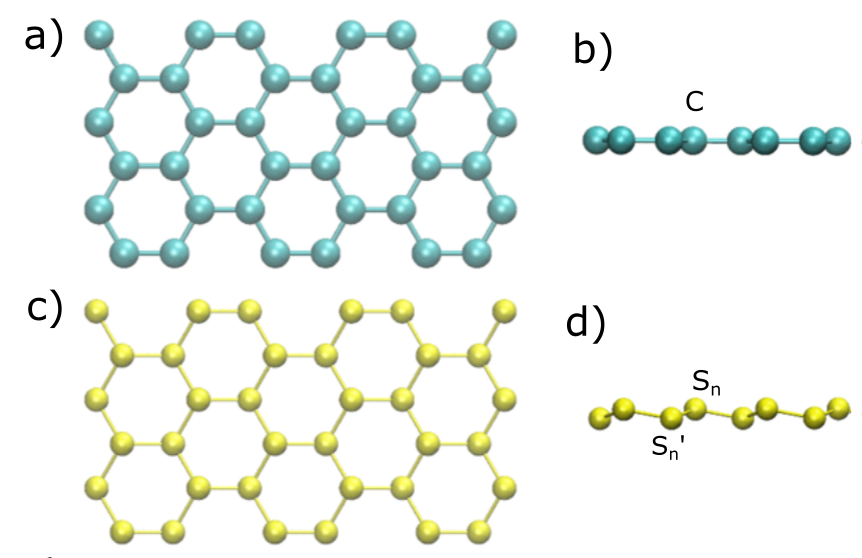}
    \caption{(a) Vista superior de la estructura de tipo panal de abejas o \textit{honeycomb} del grafeno. (b) Vista lateral de la estrucutra del grafeno. (c) Vista superior de la estructura del “estaneno” libre. (d) Vista lateral de la estructura del “estaneno” libre. La misma no resulta plana, sino que presenta corrugación (\textit{buckled}). La figura fue adaptada de \cite{figStan}.}
    \label{fig:grafeno}
\end{figure}

Desde que en 2004 Geim y otros colaboradores de la Universidad de Manchester aislaran por primera vez una monocapa de grafeno de una muestra de grafito, el interés por los cristales bidimensionales creció abruptamente, en parte porque se pensaba que fuesen inestables a temperatura finita\cite{Matthew}. El grafeno es un compuesto bidimensional a base de C que posee una estructura hexagonal de tipo panal de abejas (\textit{honeycomb}) como se muestra en la Figura \ref{fig:grafeno} (a),(b). Investigaciones reportan propiedades superlativas para este material: alta conductividad eléctrica y resistencia mecánica, baja reactividad química, entre otras\citep{Andrew}. La técnica que se empleó para la obtención de la primera monocapa de grafeno se denomina exfoliación. En ella, se extrae una monocapa con cinta adhesiva de un volumen (\textit{bulk}) que posee capas bidimensionales del compuesto a aislar (por ejemplo, del grafito).

Dadas las notables propiedades del grafeno, se ha promovido la búsqueda de nuevos materiales bidimensionales que tuviesen propiedades similares. Entre ellos, las posibles variantes: “siliceno”, “germaneno”, “estaneno”, etc., serían análogos al grafeno pero conformados exclusivamente por átomos de Si, Ge y Sn respectivamente\cite{Andrew}. Predicciones establecen que el “estaneno” sería el candidato ideal para que se manifestase el efecto Hall Cuántico de Spin o QSH (\textit{Quantum Spin Hall effect})\cite{aislante}. Este efecto aparece en los aislantes topológicos, y es el responsable de que exista un estado superficial protegido por la simetría de inversión temporal, y que permitiría la difusión de electrones por la superficie sin disipación. Esto último tendría eventuales aplicaciones en la spintrónica y en computación cuántica\cite{Andrew}. El QSH al parecer no ocurriría ni en el grafeno, ni el siliceno ni en el germaneno dado que el acomplamiento spin-órbita SOC (\textit{Spin-Orbit Coupling}) resulta ser relevante sólo para elementos pesados\cite{aislante}.

Para sintetizar el “estaneno” no podría emplearse la técnica de exfoliación ya que en la naturaleza no existe un \textit{bulk} compuesto por capas bidimensionales de Sn. Además, la predicciones indican que la estructura de tipo panal de abejas del “estaneno” libre no sería plana, más bien estaría corrugada o \textit{buckled} (ver Figura \ref{fig:grafeno} (c),(d)). Una técnica alternativa es el crecimiento epitaxial, en donde se vapora Sn sobre una superficie de estructura hexagonal con un parámetro de red conveniente en condiciones de ultra alto vacío. El primer reporte de síntesis de “estaneno” apareció en 2015\cite{Zhu}. En el año 2018, Junji Juhara \textit{et al}\cite{Yuhara} reportaron la primera síntesis de “estaneno” con una estrucutra plana, es decir, sin corrugación, al haber evaporado Sn en Ag(111). 

En el presente trabajo se buscará replicar las mediciones hechas por Junji Juhara \textit{et al} para verificar la reproducibilidad de los resultados.
\section{\label{sec:met}Método experimental}

Los experimentos se realizaron utilizando una muestra de Ag(111) sobre la cual se evaporó Sn, y una muestra de Au(111). Las técnicas empleadas para las mediciones fueron: difracción de electrones de baja energía (LEED), microscopía de efecto túnel (STM), espectroscopía de fotoelectrones emitidos por Rayos X (XPS) y espectroscopía de fotoelectrones resuelta en ángulo (ARPES/UPS). En el laboratorio estas  técnicas no se encontraban todas juntas en la misma cámara de ultra-alto vacío (UAV), más bien separadas en dos equipos independientes: por un lado el LEED y el STM (equipo LEED-STM), y por el otro, el XPS y el ARPES/UPS (equipos XPS-UPS). El ultra-alto vacío fue necesario para mantener al mínimo la contaminacíon de las muestras durante el día de medición.

El equipo STM/LEED contaba con una cámara principal de ultra alto vacío (con una presión base de 2-3$\cdot10^{-10}$ Torr) y una cámara auxiliar para la introduccion de muestras. El sistema de vacío disponía de una bomba iónica instalada en la cámara principal y una bomba turbomolecular instalada en la pre-cámara. Cuando se bombardeaba con Ar$^+$ para realizar el \textit{sputtering} la bomba iónica se aislaba y sólo se utilizaba la bomba turbomolecular para el bombeo de las dos cámaras.

La cámara principal del equipo LEED/STM estba equipada, además del microscopio STM y de la óptica de LEED, con un cañón de iones de Ar$^+$,  un manipulador que permitía llevar a cabo el recocido de las muestras y un evaporador de Sn.

Por otro lado, el equipo XPS/UPS también disponía de una cámara principal y de una cámara auxiliar para la introducción de muestras, las cuales poseían sendas bombas turbomoleculares instaladas. 
Las presiones de base de dichas cámaras fueron de 2-3$\cdot10^{-10}$ y 8$\cdot10^{-8}$ Torr, respectivamente. 

La cámara principal del equipo XPS/UPS contaba con un analizador hemisférico de energía cinética, una fuente de rayos X (que emitía radiación de 1486,6 eV debido a la línea $K_{\alpha}$ del Al) acoplada a un monocromador, una lámpara de descarga de Helio (para realizar los experimentos de UPS) y un cañon de Ar$^+$. Mientras que en este equipo no se disponía de una óptica de LEED. El manipulador de la cámara principal tenía lugar para 4 muestras. Una de dichas posiciones ofrecía la posibilidad  para llevar a cabo el calentamiento de las mismas. Además, podían enfriarse hasta aproximadamente 90 K mediante un sistema de flujo continuo de Nitrógeno líquido. 

La evaporación de Sn se realizó en la cámara auxiliar, para lo cual se instaló un evaporador de Sn hecho ad-hoc. El evaporador consistía de un crisol de nitruro de boro,  calentado resistivamente por un filamento de Tugsteno. Para evaporar Sn se necesitaba alcanzar temperaturas de alrededor de 700$^{\circ}$C.

\subsection{Preparación de la muestra}

Antes de cada medición la superficie de la muestra de Ag(111) se preparaba mediante ciclos de \textit{ sputtering} y \textit{anneling}.

La técnica del \textit{sputtering} consiste en el bombardeo de la muestra con átomos de Ar$^+$ con el fin de eliminar los contaminantes adsorbidos en la superficie. En ambos equipos la tensión aplicada entre la muestra y la fuente de iones fue de 1,5 kV, siendo la corriente de iones de 12 $\mu$A (P$_{Ar}$=1,2$\cdot10^{-7}$ Torr) y 1 $\mu$A (P$_{Ar}$=2,5$\cdot10^{-5}$ Torr) en los equipos XPS/UPS y LEED/STM respectivamente. P$_{Ar}$ indica la presión parcial de Ar$^+$.

La técnica del \textit{anneling} (o recocido) consiste en el calentamiento de la muestra con el fin de brindar movilidad a los átomos, de manera tal de alcanzar una nueva configuración con una estructura más ordenada. Típicamente se necesitan temperaturas del orden de 500$^{\circ}$C para lograr superficies ordenadas en el caso de metales nobles. En el equipo XPS-UPS el calentamiento se realizaba a través de un filamento que se encontraba por debajo de la muestra en la cámara principal. El sistema de calentamiento poseía dos modos de operación: radiativo (modo \textit{filament}) y bombardeo electrónico (modo \textit{emission}). En el primero solo se hacía pasar corriente por el filamento mientras que en el segundo también se aplicaba una diferencia de potencial entre filamanto y muestra, de forma tal que los electrones emitidos por el mismo fuesen acelerados hacia la muestra. La determinación de la temperatura de la muestra era aproximada dado que la termocupla utilizada no se encontraba en el portamuetras, sino en el manipulador. Además, la temperatura de la muestra difería respecto la temperatura medida por la termocupla dependiendo del modo de calentamiento: por ejemplo, en el modo de bombardeo electrónico, dependía de la potencia aplicada (a mayor potencia aplicada mayor resultaba la diferencia de temperarturas). En el caso del equipo LEED/STM el sistema de calentamiento era más controlado. Se utilizó sólo calentamiento de tipo radiativo. En ambos equipos durante el recocido la presión era del orden de 10$^{-9}$ Torr.    

La deposición de Sn se realizó a dos temperaturas distintas según la cantidad de Sn ya depositada, con el fin de replicar la metodología empleada en\cite{Yuhara}: a 200 $^{\circ}$C si la cantidad depositada de Sn era menor a 1/3 de MC y a 150 $^{\circ}$C si la misma era mayor o igual a 1/3 de MC. Se entiende por 1 MC (1 monocapa) a una cantidad de Sn tal que la relación de Sn/Ag en la primera capa de la Ag(111) es uno a uno. 

Antes de realizar cada evaporación, se iba aumentando progresivamente la potencia del evaporador hasta llegar a la potencia de emisión (aproximadamente 50 W en el equipo LEED-STM y 70W en el equipo XPS-UPS). Luego se dejaba al evaporador en tal valor de potencia, y tras 15 minutos de precalentamiento, la muestra se ponía en dirección al evaporador para así iniciar con la evaporación. En el equipo XPS-UPS la presión durante el proceso fue del orden de $10^{-8}$ Torr, mientras que en el equipo LEED-STM fue de $10^{-9}$ Torr.

Uno de los problemas a resolver en este tipo de experimentos es la calibración de los evaporadores. El método usual consiste en utilizar como referencia una estructura de superficie con cubrimiento conocido. Para identificar la fase de referencia se utiliza la técnica de LEED. 

El evaporador del equipo LEED-STM se encontraba calibrado por el personal del Laboratorio de Superficies, con una tasa de emisión de aproximadamente 0,035 MC/min. En cambio, la calibración del evaporador instalado en el equipo XPS/UPS resultó menos precisa debido a la falta de una óptica de LEED.  Se procedió de la siguiente manera: se prepararon muestras Sn/Ag(111) con una dosis conocida de Sn en el equipo LEED/STM, se las extrajo (estando en contacto con el aire) y finalmente se las introdujo en el equipo de XPS/UPS. Luego, estas muestras fueron tomadas como patrones para calibrar el evaporador de Sn.  

\subsection{XPS}
\begin{figure}[ht]
    \centering
    \includegraphics[width=\linewidth]{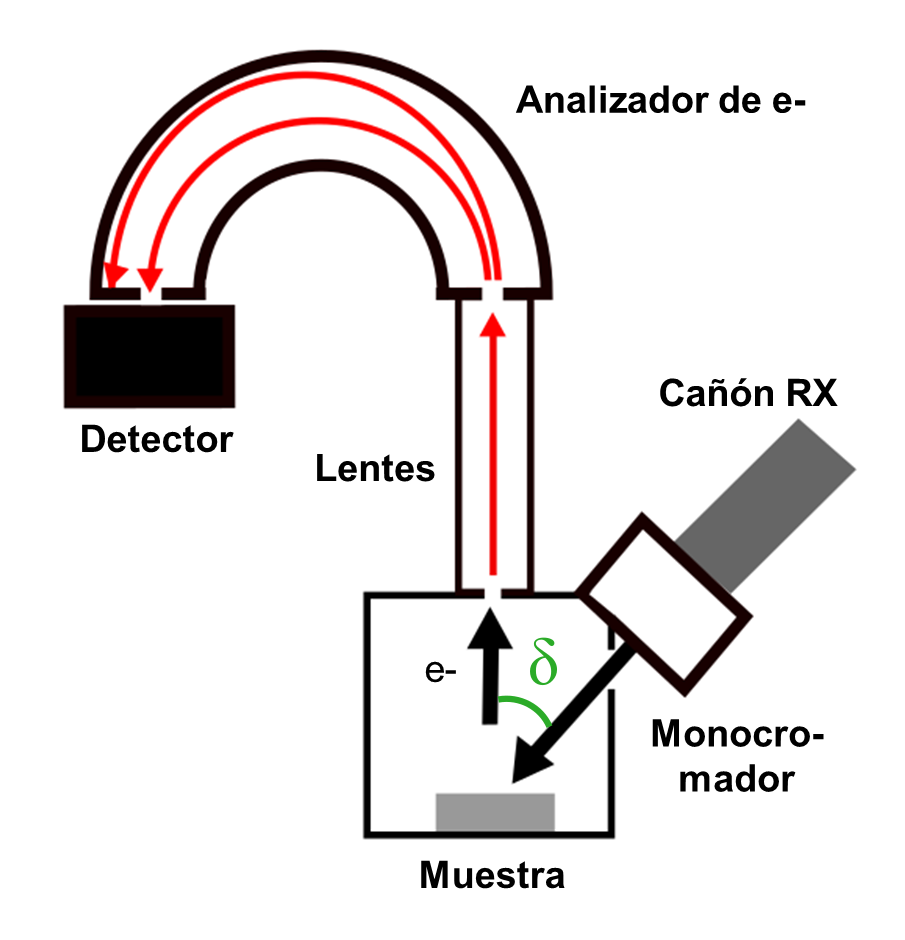}
    \caption{Esquema experimental del equipo de XPS. Un cañón de Rayos X emite radiación monocromática que incide en la muestra. Los electrones fotoemitidos son enfocados y frenados por un conjunto de lentes electroestáticas. Un analizador de electrones sólo deja pasar a aquellos que posean una determinada energía cinética. Por último, un detector cuenta el número de electrones que logran atraversar al analizador. El ángulo entre el haz incidente y la dirección de observación $\delta\approx 54^{\circ}$.}
    \label{fig:xps}
\end{figure}

La técnica de espectroscopía de fotoelectrones emitidos por Rayos X o XPS (\textit{X-Ray Photoelectron Spectroscopy}) en esencia consiste en la irradiación de la muestra bajo estudio con un haz monocromático de  Rayos X, y en el posterior análisis energético de los electrones emitidos. En la Figura \ref{fig:xps} se muestra un esquema del detector. 

El cañón de Rayos X (RX) emite fotones asociados a la línea $K_{\alpha}$ del Al con una energía de 1486,6 eV, cuyo ancho de línea es reducido por un monocromador. Los fotones interaccionan con los electrones en los estados internos de la muestra o \textit{Core levels} y aquellos que son emitidos elásticamente adquieren una energía cinética $E_k$ tal que
\begin{equation}
E_k=h\nu-E_b-\phi,
\label{eq:ecin}
\end{equation}
donde $h\nu$=1486,6 eV, $E_b$ es la energía de ligadura medida respecto del nivel de Fermi y $\phi$ es la función trabajo de la muestra. Luego los electrones son enfocados y frenados por un conjunto de lentes electroestáticas hasta ingresar a un analizador hemiesférico. El frenamiento ocurre pues las mediciones fueron realizadas con energía de paso constante. La energía de paso es la energía a la cual los electrones entrantes al analizador pueden atraversarlo completamente, dado que las dos semiesferas del mismo se encuentran a una diferencia de potencial constante fijada por tal energía. Para realizar la medición de un espectro, un detector registra el número de cuentas al ir variando la energía del frenado. La principal ventaja de utilizar el modo de medición con energía de paso constante, se debe a que la resolución energética a lo largo de todo un espectro resulta aproximadamente constante.

La técnica de XPS además de ser sensible a la distintas especies químicas presentes en la muestra, también resulta ser sensible a su composición relativa. La intensidad $I_{X,nl}$ del pico asociado al nivel $nl$ de la especie química X viene dada por:
\begin{equation}
I^{X,nl} =AF\Delta \Omega  N_x\frac{d\sigma_{X,nl}}{d\Omega}\lambda(E^{X,nl}_k)T(E^{X,nl}_k),
\label{eq:int}
\end{equation}
en donde $A$ es el área de la muestra de donde los foloelectrones son colectados, $F$ es el flujo de fotones, $\Delta \Omega$ es el ángulo sólido de aceptación del detector, $N_x$ es la densidad de átomos de la especie $X$, $\frac{d\sigma_{X,nl}}{d\Omega}$ es la sección eficaz diferencial de fotoemisión, $\lambda$ es el camino libre medio de los electrones y $T$ la eficiencia del analizador. Además $E^{X,nl}_k$ se refiere a la energía cinética de los fotoelectrones asociados al nivel $nl$ de la especie química $X$. Por otro lado, $\frac{d\sigma_{X,nl}}{d\Omega}$ se relaciona con la sección eficaz total $\sigma_{X,nl}(E^{X,nl}_k)$ para un haz de fotones no polarizados según:
\begin{equation}
\begin{split}
\frac{d\sigma_{X,nl}}{d\Omega}& =\frac{\sigma_{X,nl}(E^{X,nl}_k)}{4\pi}\left(1-\frac{\beta_{nl}(E^{X,nl}_k)}{2}P_2(\cos(\delta))\right) \\
&\approx \frac{\sigma_{X,nl}(E^{X,nl}_k)}{4\pi},
\end{split}
\label{eq:intdos}
\end{equation}
en donde $\beta_{nl}$ es un parámetro de asimetría, $P_2$ es el segundo polinomio de Legendre y $\delta\approx54^{\circ}$ es el ángulo entre el haz incidente y la dirección de observación.
A partir de la Ecuaciones \ref{eq:int} y \ref{eq:intdos} puede obtenerse una expresión para la dosis $\epsilon$ de Sn en la muestra de Ag:
\begin{equation}
\epsilon=\frac{N_{Sn}}{N_{Ag}}\frac{\lambda_{Sn}T_{Sn}}{\lambda_{Ag}T_{Ag}}=\frac{I_{Sn(3d)}}{I_{Ag(3d)}}\frac{\sigma_{Ag(3d)}}{\sigma_{Sn(3d)}}\approx 0,718\frac{I_{Sn(3d)}}{I_{Ag(3d)}},
\label{eq:dosis}
\end{equation}
donde se utilizó que $\sigma_{Sn(3d)}=0,344\cdot10^6$ barn y $\sigma_{Ag(3d)}=0,247\cdot10^6$ barn para $h\nu$=1486,6 eV \cite{Yeh}.
\subsection{ARPES/UPS}
La técnica de espectroscopía de fotoelectrones resuelta en ángulo o ARPES (\textit{Angle-resolved photoemission spectroscopy}) es un método que permite determinar las bandas electrónicas de los sólidos. Al igual que en la técnica de XPS, se irradia la muestra bajo estudio con un haz monocromático, en este caso, con luz ultravioleta (de ahí UPS, por \textit{ultraviolet}) siendo la energía de los fotones de 21,2 eV debido a la emisión de la línea 1$\alpha$ de la lámpara de He. Posteriormente, a partir del análisis de la energía cinética de los electrones fotoemitidos, se pueden generar espectros para distintos ángulos de incidencia de la radiación $\theta$ respecto de la dirección normal de la muestra.

El proceso completo de la fotoemisión se compone de tres partes. Primero, los electrones son excitados por la radiación cumpliendo con la conservación del momento total:
\begin{equation}
\hbar\vec{k}_f=\hbar\vec{k}_i+\hbar\vec{k}_{\gamma}+\hbar\vec{G}\approx\hbar\vec{k}_i+\hbar\vec{G},
\end{equation}
siendo $\vec{k}_i$ y $\vec{k}_f$ los vectores de onda del estado inicial y excitado respectivamente, $\vec{k}_{\gamma}$ el vector de onda del fotón incidente y $\vec{G}$ un vector de la red recíproca. Dado que la energía de los fotones incidentes era de 21,2 eV, $\vec{k}_{\gamma}\approx$ 0,01Å$^{-1}$ resulta ser despreciable frente al tamaño típico de la primera zona de Brillouin. Entonces, como el vector de onda del estado excitado difiere sólo en un vector de la red recíproca respecto del vector de onda inicial, se dice que la transición es “vertical”; Segundo, los fotoelectrones en su camino hacia la superficie pueden sufrir procesos inelásticos que degraden su energía. Los únicos que llevan información del estado inicial son aquellos que no han experimentado procesos inelásticos; Tercero, si la superficie forma un arreglo bidimensional de largo alcance, la transmisión de los fotoelectrones conserva la componente paralela a la superficie de sus vectores de onda, es decir, $\vec{k}_{f\parallel}=\vec{k}_{\parallel}$ donde $\vec{k}$ es el vector de onda de los fotoelectrones en el vacío. 

En conclusión, dado que la componente paralela a la superficie del vector de onda se conserva y que la fotoexitación es una transición vertical, se puede conocer $\vec{k}_{i\parallel}$ de un nivel del sólido a través de un experimento de fotoemisión resuelta en ángulo. En el caso de un estado electrónico de superficie, esto alcanzaría para determinar completamente el vector de onda del estado inicial.

En términos cuantitativos, como la  energía cinética $E_k$ de una partícula libre se relaciona con su número de onda $\vec{k}$ de forma tal que $E_k=\frac{\hbar^2k^2}{2m}$ y como $k=k_{\parallel}\sin(\theta)$, donde $k_{\parallel}$ es la proyección del número de onda en la superficie (que coincide con $k_{i\parallel}$), resulta que:
\begin{equation}
k_{i\parallel}=\frac{\sqrt{2m}}{\hbar}\sin(\theta)\sqrt{E_k}=0,512\:\sin(\theta)\sqrt{E_k(\text{eV})}\:\:\left[\text{Å}^{-1}\right].
\label{eq:kpar}
\end{equation}
Luego, resulta que para los estados de superficie que se encuentren en la banda de valencia de la muestra, la técnica de ARPES/UPS permite obtener el valor de $k_{i\parallel}$ para un dado ángulo $\theta$, y por ende, la relación de dispersión del estado ($E_b$ en función de $k_{i\parallel}$).
\subsection{LEED}
\begin{figure}[ht]
    \centering
    \includegraphics[width=\linewidth]{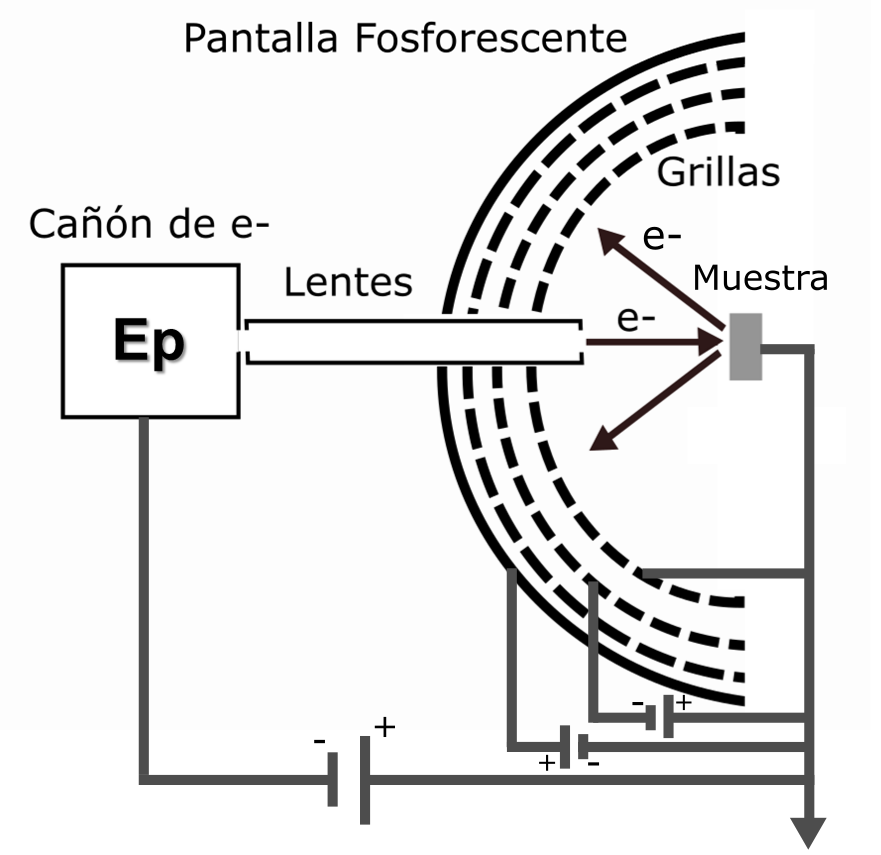}
    \caption{Esquema experimental del equipo de LEED. Un cañón de electrones emite un haz de electrones monoenergéticos de energía $E_p$. El mismo es enfocado por un conjunto de lentes para que incida normalmente respecto de la superficie de la muestra. Los electrones son difractados y un conjunto de grillas filtra solo a aquellos que lo hayan hecho elásticamente. Por último, éstos impactan contra una pantalla fosforescente produciendo un patrón característico de la muestra.}
    \label{fig:leed_equipo}
\end{figure}

La técnica de difracción de electrones de baja energía o LEED (Low-Electron Energy Diffraction) consiste en el bombardeo de la muestra en estudio con un haz monoenergético de electrones normal a su superficie, que produce luminosidad en una pantalla fosforescente debido a los electrones difractados elásticamente. En la Figura \ref{fig:leed_equipo} se muestra un esquema del equipo. Para garantizar que sólo los electrones que hayan sido difractados elásticamente impactasen la pantalla fluorescente, el equipo disponía de un conjunto de grillas que frenaban a todos los electrones que tuviesen una energía cinética menor que la energía cinética de los electrones incidentes. Además, si la superficie forma una arreglo bidimensional con orden de largo alcance,  los electrones dispersados deben cumplir Ley de Bragg en 2 dimensiones. En tal caso se cumplen las relaciones:
\begin{equation}
\begin{split}
\begin{cases}\vec{k}_{f\parallel} & =\vec{k}_{i\parallel}+\vec{G}_{\parallel},  \\
|\vec{k}_{f}| & =|\vec{k}_{i}|,
\end{cases}
\end{split}
\end{equation}
donde $\vec{k}_{i\parallel}$ y $\vec{k}_{f\parallel}$ son las componentes superficiales de los números de onda asociados a los electrones incidentes y refractados respesctivamente, y $\vec{G}_{\parallel}$ un vector de la red recíproca de la superficie. El patrón de difracción resultante constituye una réplica a escala de la red recíproca de la superficie. Por lo tanto, a partir de la técnica del LEED es posible inferir la estructura cristalina subyacente de la superficie de la muestra.

\subsection{STM}

El microscopio de efecto túnel o STM (\textit{Scanning tunneling microscope}) permite conocer la topología de la superficie de una muestra conductora a nivel atómico. Para ello, el microscopio posee una punta metálica que se acerca a la muestra. Entre ambas, se aplica una tensión $V$ que produce una corriente $I$ por efecto túnel. Dado que $I\propto e^{-az}$ siendo $a$ una constante y $z$ la altura de la punta metálica respecto de la muestra, la corriente túnel resulta ser muy sensible a pequeñas variaciones en la altura. Por lo tanto, el STM permite obtener información a escala atómica de la superficie. Para realizar las mediciones se utilizó el modo de corriente constante en el cual al desplazarse en el plano xy, la punta metálica va regulando la altura con el fin de mantener constante la corriente establecida.

\section{\label{sec:level1}Resultados y discusión}

\subsection{Análisis de la Estructura Cristalina}
\begin{figure}[ht]
    \centering
    \includegraphics[width=\linewidth]{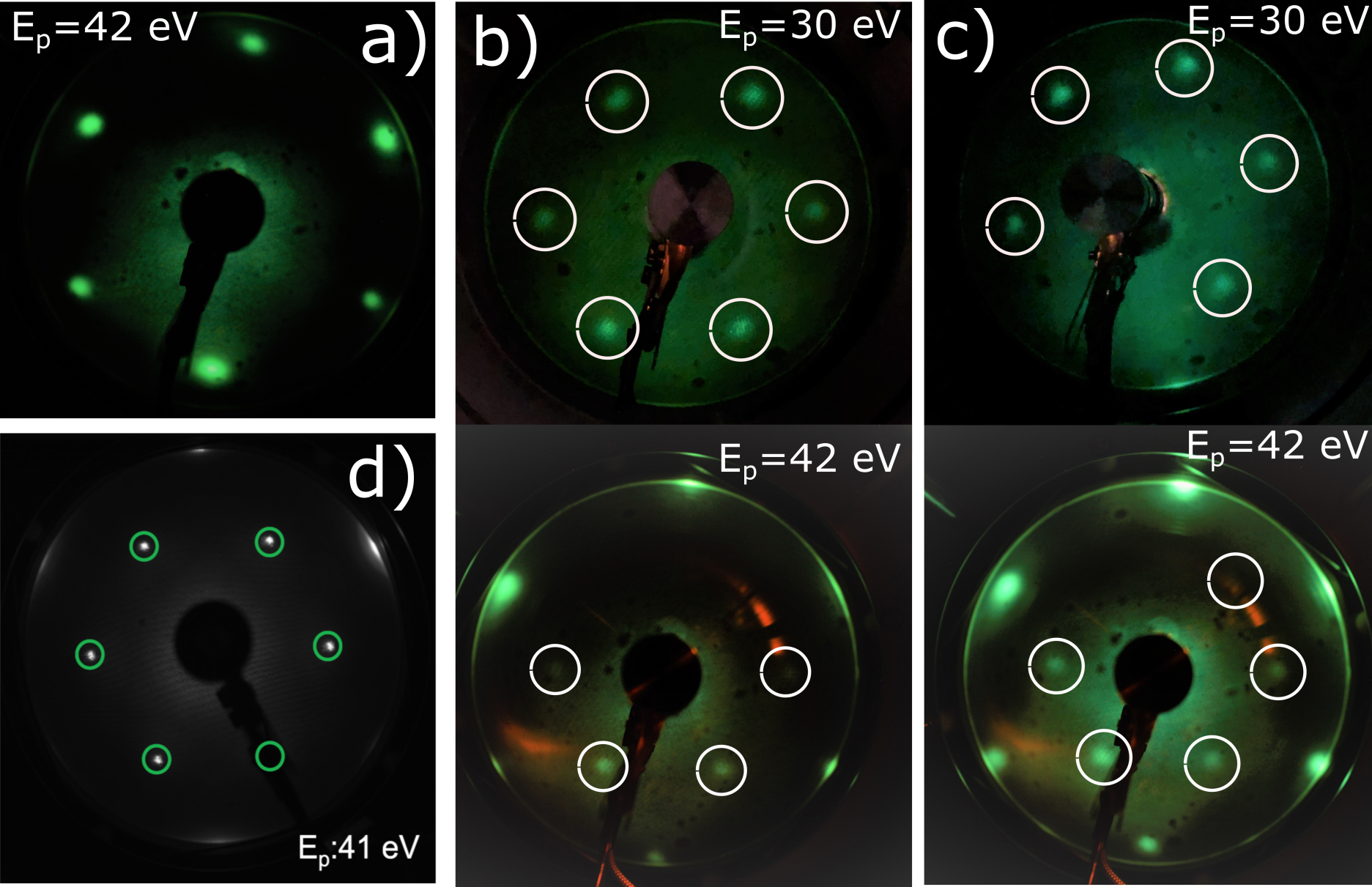}
    \caption{Patrones de LEED de Ag(111) y de Sn/Ag(111). En todos lo casos $E_p$ indica la energía cinética de los electrones incidentes sobre la muestra. (a) Patrón hexagonal consistente con una estructura $1\times 1$ de Ag(111) limpia. (b) Patrón hexagonal (señalado por los círculos) consistente con una estructura ($\sqrt{3}\times \sqrt{3}$)R30$^{\circ}$, para 1/3 de MC de Sn. (c) Patrón hexagonal (señalado por los círculos) para (1/3+0,4) de MC de Sn. (d) Patrón para 1/3 de MC de Sn según \cite{Yuhara}. }
    \label{fig:leed}
\end{figure}

\begin{figure}[ht]
    \centering
    \includegraphics[width=0.9\linewidth]{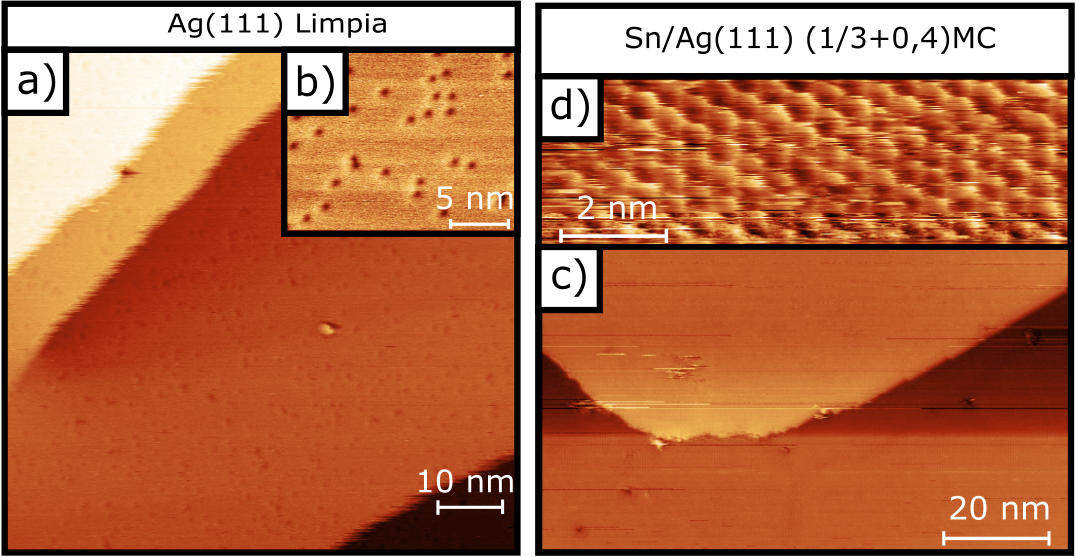}
    \caption{Imágenes obtenidas con el STM. En (a) y (b) la corriente túnel fue de 2 nA, mientras que la tensión fue de -0,5 V y de -0,1 V respectivamente. En (c) y (d) la tensión fue de +0,5 V y la corriente túnel de 1 nA.}
    \label{fig:stm}
\end{figure}

Utilizando la técnica de LEED se obtuvieron patrones de difracción para distinas cantidades de Sn en Ag(111) (ver Figura \ref{fig:leed}). Para la muestra de Ag(111) pura puede observarse un patrón de difracción hexagonal correspondiente a una estructura $1\times1$. El hecho de que los puntos del patrón de LEED tuviesen cierto grosor indica que el orden de largo alcance no es suficiente para definir los puntos como corresponde a superficies de Ag(111) de buena calidad\cite{Yuhara}. En la Figura \ref{fig:stm}(a) se observa la presencia de terrazas en una imagen obtenida con el STM. Aún así, también pueden observarse regiones planas. Por otro lado, en la Figura \ref{fig:stm}(b) se observan puntos que podrían estar asociados a átomos de Sn embebidos en la primera capa de la muestra.

\begin{figure}[ht]
    \centering
    \includegraphics[width=\linewidth]{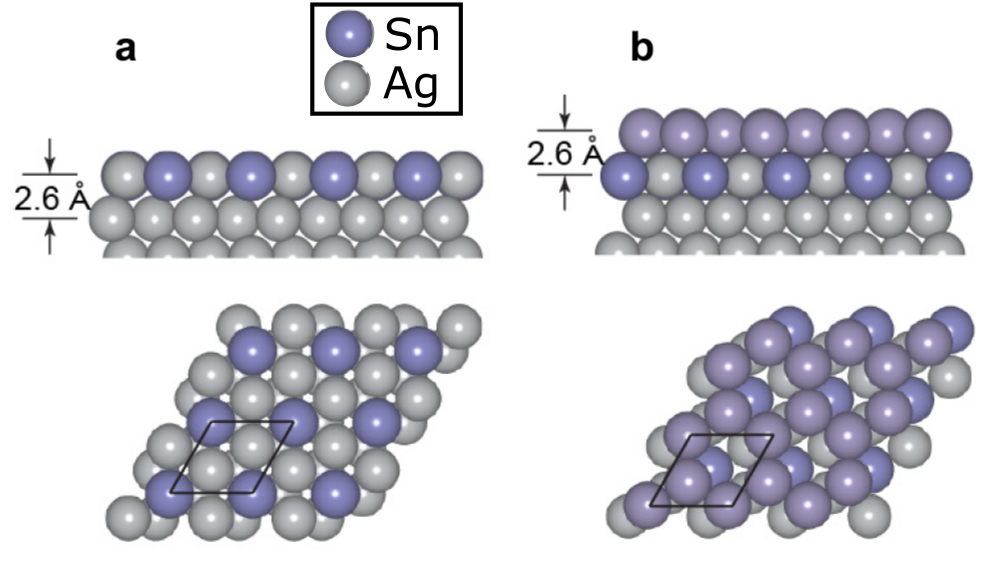}
    \caption{Modelos estructurales de la aleación de superficie Ag$_2$Sn para 1/3MC de Sn (a) y del “estaneno” sobre la aleación de Ag$_2$Sn para (1/3+2/3)MC de Sn (b). La imagen fue tomada de \cite{Yuhara}.}
    \label{fig:estrucutra}
\end{figure}

Al evaporar  1/3 de MC de Sn, además del patrón $1\times1$ asociado a la Ag, aparece un nuevo patrón hexagonal más pequeño y rotado en 30$^{\circ}$. El mismo está asociado a un estructura de tipo ($\sqrt{3}\times \sqrt{3}$)R30$^{\circ}$ para los átomos de Sn. Esto último es consistente con la ya conocida estructura de la aleación de superficie Ag$_2$Sn \cite{Yuhara}\cite{Jacek}. La disposición de los átomos de Sn que adoptarían en la Ag(111) puede observarse en la 
Figura \ref{fig:estrucutra}(a). Por otro lado, al evaporar (1/3+0.4) de MC de Sn el patrón de difracción continua teniendo la misma estructura con respecto al patrón de 1/3 de MC de Sn. Según \citep{Yuhara}, la coincidencia podría explicarse por un crecimiento por capas, en donde el “estaneno” se formaría sobre la aleación de Ag$_2$Sn (ver Figura \ref{fig:estrucutra}(b)). En la Figura \ref{fig:stm} (c) aún pueden observarse regiones planas tras la evaporación de (1/3+0.4) de MC de Sn, mientras que en (d) se observa la superficie con resolución atómica. La imagen  muestra una estructura con el aspecto y las dimensiones  de la reconstrucción ($\sqrt{3}\times \sqrt{3}$)R30$^{\circ}$\cite{Jacek}. Por otro lado, para formar completamente la capa de “estaneno” se necesitaría depositar (1/3+0,66)MC de Sn, por lo cual  la superficie preparada podría tener zonas con la estructura expuesta\cite{Yuhara}. 

\subsection{Análisis de la composición química}

\begin{figure}[ht]
    \centering
    \includegraphics[width=0.9\linewidth]{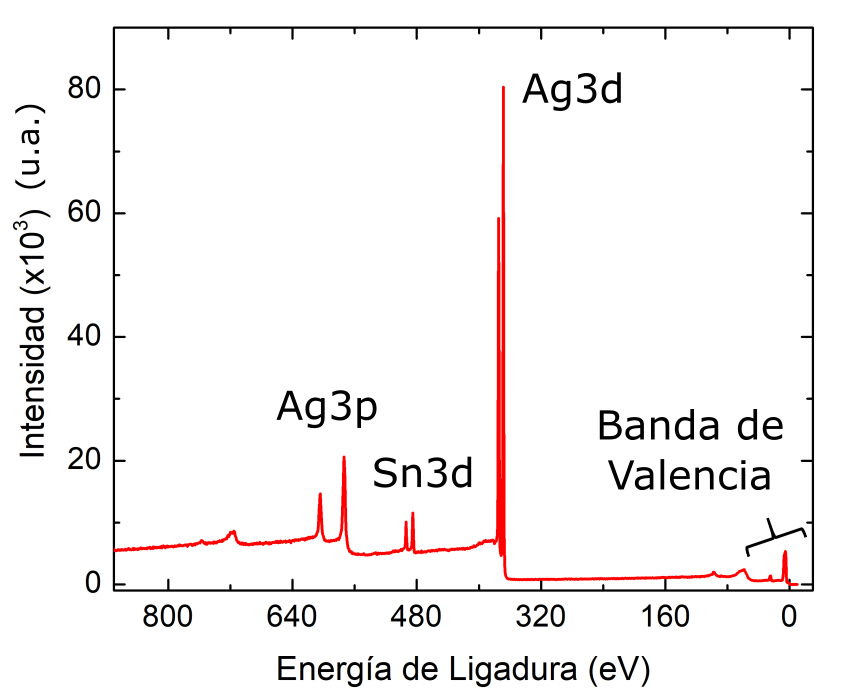}
    \caption{Espectro de XPS de la muestra de Sn/Ag(111). Se observan picos asociados a algunos niveles internos (\textit{Core Levels}) y a la Banda de Valencia de los átomos presentes en la muestra.}
    \label{fig:survey}
\end{figure}

Para estudiar la composición química de la muestra se utilizó la técnica de XPS. Un ejemplo de un espectro se muestra en la Figura \ref{fig:survey}.

En el espectro se observa un conjunto de picos que están asociados a los niveles internos o \textit{Core Levels} de los átomos presentes en la muestra. Algunos de ellos se encuentran desdoblados debido a la interacción spin-órbita. También puede notarse la presencia de un fondo cuyo origen se atribuye a electrones que fueron dispersados inelásticamente, es decir, a aquellos que interactuaron y perdieron parte de su energía cinética antes de abandonar la muestra. 

\begin{figure}[ht]
    \centering
    \includegraphics[width=\linewidth]{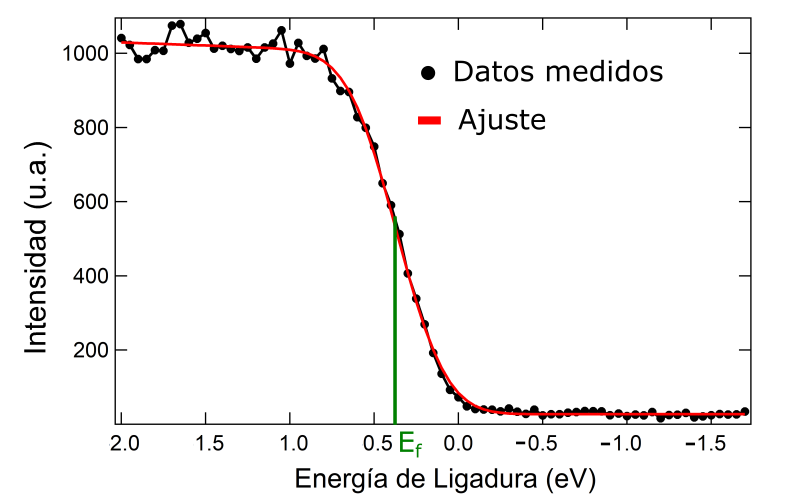}
    \caption{Ajuste del borde de Fermi. La línea vertical indica la posición de la energía de Fermi.}
    \label{fig:fitfermi}
\end{figure}

Por otro lado, en un entorno del cero de energía se observa que la intensidad del espectro decrece casi abruptamente para energías negativas. Esta caída abrupta (ver Figura \ref{fig:fitfermi}) se  debe a aque los electrones siguen la estadística de Fermi-Dirac. Por lo tanto, el borde de la Banda de Valencia tiene la forma funcional dada por la distribución de Fermi convolucionada con una Gaussiana. Ésta última representa la resolución finita en energía del sistema compuesto por la fuente de fotones y el analizador. 

Los valores obtenidos para el ancho Gaussiano a mitad altura FWHM fueron del orden de 0,5 eV a temperatura ambiente. Dado que la energía de fermi $E_F$ es una propiedad intrínseca de la muestra, el valor obtenido en cada espectro se utilizó para calibrarlos, es decir, para cada uno de ellos se redefinió el cero de energía de forma tal que $E_F\equiv 0$.


Para relacionar la cantidad de Sn depositado en las muestras preparadas en el equipo XPS-UPS con las muestras patrones traidas del equipo LEED-STM, se utilizó el cociente de intensidades $\epsilon$ definido en Ecuación \ref{eq:dosis}, según la cual $\epsilon \approx 0,718\frac{I_{Sn}}{I_{Ag}}$. Se emplearon como intensidades las áreas de los picos del Sn3d5/2 y de la Ag3d5/2.

\begin{figure}[ht]
    \centering
    \includegraphics[width=0.9\linewidth]{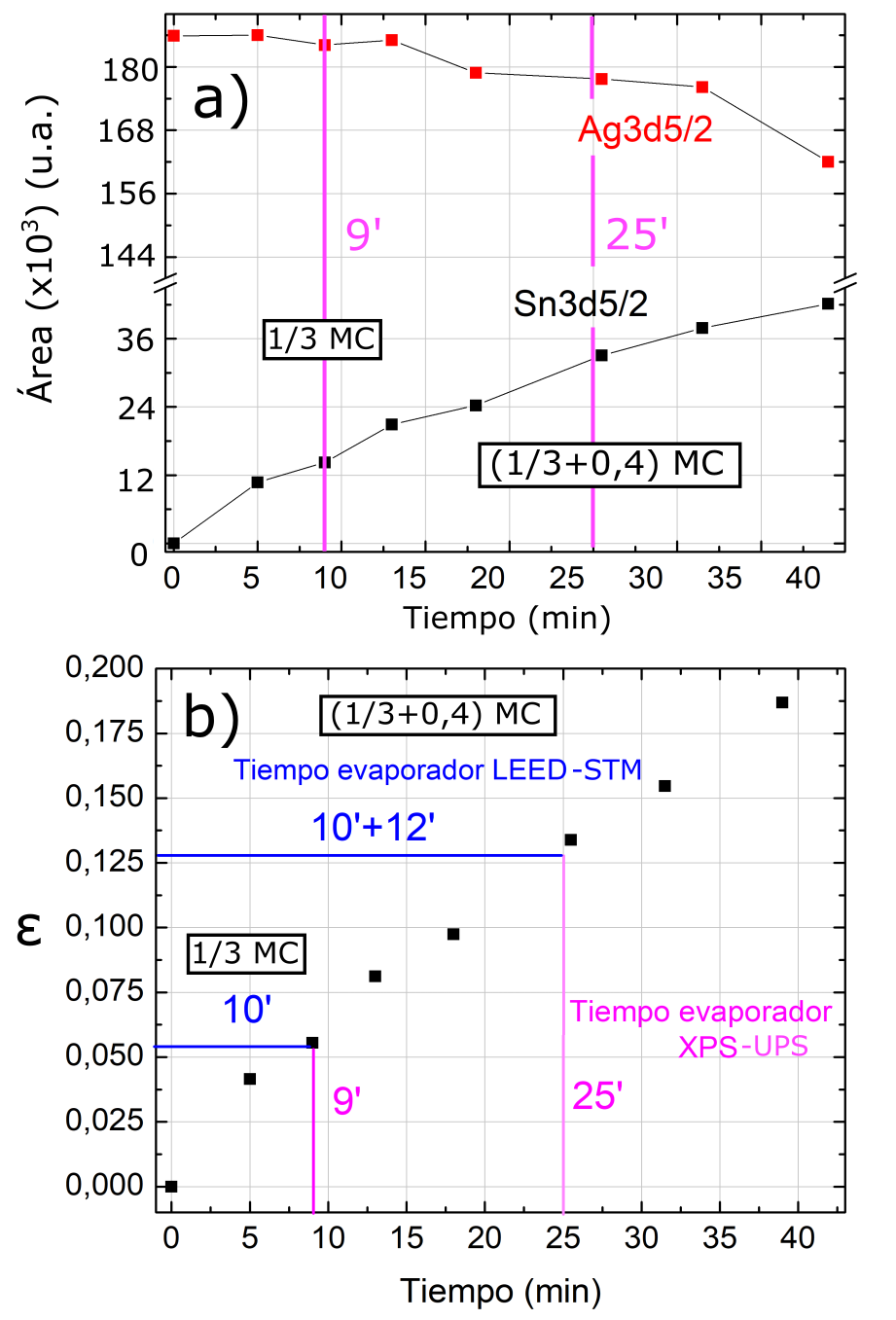}
    \caption{(a) Área de los picos en función del tiempo neto de evaporación. (b) Cociente $\epsilon$ de Sn en la muestra de Ag(111) en función del tiempo neto de evaporación. Las líneas horizontales indican las dosis de dos muestras cuya preparación se realizó en el equipo LEED-STM, habiendose depositado 1/3 y (1/3+0,4) de MC de Sn respectivamente.}
    \label{fig:area_tiempo}
\end{figure}

En la Figura \ref{fig:area_tiempo}(a) se muestra el gráfico del área de los picos de Sn3d5/2 y de Ag3d5/2 en función del tiempo neto de evaporación, mientras que en la Figura \ref{fig:area_tiempo}(b) se muestra la dosis de Sn en la muestra en función del tiempo neto de evaporación. Las líneas horizontales en la Figura \ref{fig:area_tiempo}(b) indican los cocientes $\epsilon$ correspondientes a las muestras patrón. De esta forma, dado que la tasa de evaporación en el equipo LEED-STM era conocida, fue posible calibrar el evaporador del equipo XPS-UPS respecto del anterior. La Figura \ref{fig:area_tiempo}(a) muestra un crecimiento continuo de la intensidad del pico del Sn3d5/2  acompañado por un decrecimiento de la intensidad del pico de  Ag3d5/2. Alrededor de los 9 minutos netos de evaporación (aproximadamente 1/3MC de Sn) se habría completado la aleación superficial de Ag$_2$Sn, sin embargo, no se observa ningún cambio en la curva del Sn que indique la aparición de una nueva fase.

\begin{figure}[ht]
    \centering
    \includegraphics[width=\linewidth]{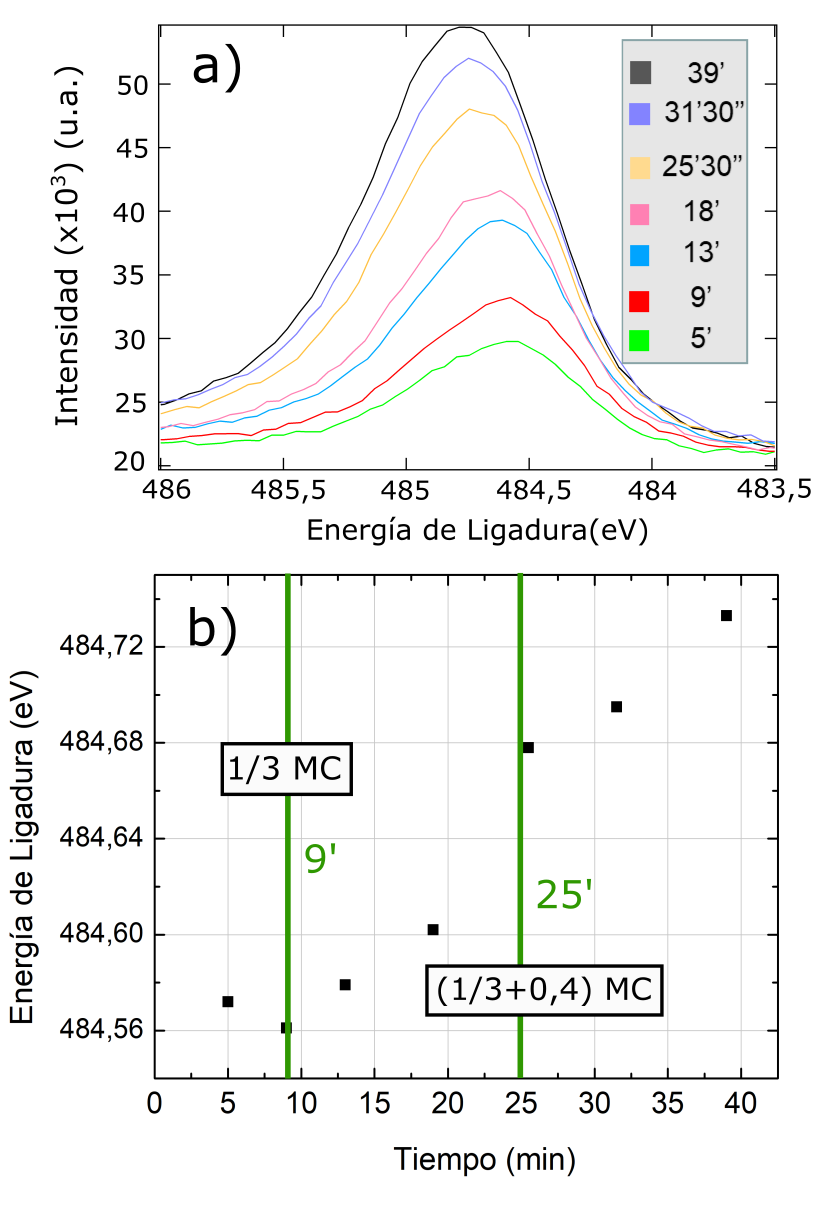}
    \caption{(a) Picos del Sn3d5/2 para distintos tiempos netos de evaporación. (b) Posición de los picos en función del tiempo neto de evaporación. El máximo corrimiento resultó ser del orden de 0,2 eV.}
    \label{fig:corri1}
\end{figure}

Donde si se observa un claro cambio es en la forma de línea de los picos Sn3d5/2. Como se puede ver en la Figura \ref{fig:corri1}(a), los picos se desplazan hacia mayores energías de ligadura y se ensanchan a medida que aumenta la dosis de Sn. En la Figura \ref{fig:corri1}(b) se muestra la posición de los picos del Sn3d5/2 en función del tiempo de evaporación neto. El corrimiento de los picos del orden de 0,2 eV es consistente con las mediciones hechas por \cite{Yuhara}. El mismo podría explicarse debido a la aparición paulatina de átomos de Sn no equivalentes que estarían asociados a la formación de la fase del “estaneno”.

\begin{figure}[ht]
    \centering
    \includegraphics[width=\linewidth]{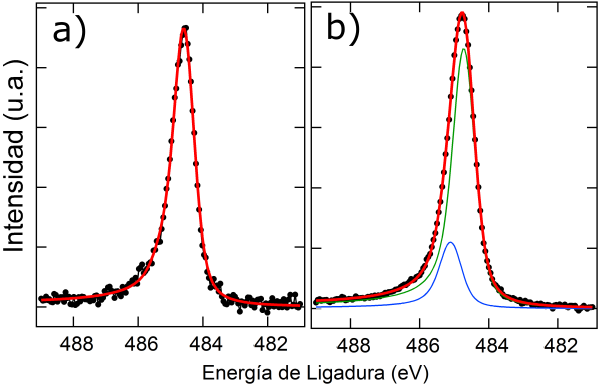}
    \caption{(En rojo) Ajustes del pico del Sn3d5/2 para 5 (a) y 39 (b) minutos netos de evaporación. En (a) se ajustó una sola componente de tipo Doniach-Sunjic, mientras que en (b) se ajustaron dos (en azul y verde).}
    \label{fig:corri2}
\end{figure}

La forma de línea de los picos de XPS es asimétrica, con una cola hacia energías de ligaduras más altas. Esta asimetría surge debido a que en compuestos metálicos, no existe un único proceso que conlleve a la emisión de los fotoelectrones. Los ajustes de tipo Doniac-Sunjic\cite{Doniach}, que tienen en cuenta un contribución lorentziana (debido a la vida media de los niveles electrónicos) asimétrica, fueron utilizados para entender más en detalle la forma de línea. Al igual que al analizar el borde de Fermi, los ajuste utilizados tuvieron en cuenta una contribución Gaussiana debido a la resolución intrínseca del equipo de medición. El espectro de la Figura \ref{fig:corri2}(a) se corresponde con un tiempo de evaporación neto de 5 minutos, por tanto, todos los átomos de Sn serían ser aproximadamente equivalentes, y el pico del Sn3d5/2 debería tener una única componente. Bajo esta suposición y fijando  el ancho Gaussiano obtenido del ajuste del nivel de Fermi de 0,58 eV, se obtuvo  una asimetría de 0,07 eV, un ancho Lorentziano de 0,35 eV y una posición de 484,56 eV. 

El ajuste de el espectro de la Figura \ref{fig:corri2}(b) que se corresponde con un tiempo de evaporación neto de 39 minutos, se realizó manteniendo los valores de asimetría, anchos Gaussianos y Lorentzianos obtenidos para el caso del ajuste la Figura 12(a). Fue necesario incluir una segunda componente para obtener un ajuste más acorde. Las posición del pico de mayor intensidad fue de 484,7 eV y el del restante 485,02 eV, siendo 0,32 eV la separación entre ellos. Este valor resultó ser el doble que el reportado por \cite{Yuhara} para la separación de las dos componentes del pico Sn4d5/2.

\begin{figure}[ht]
    \centering
    \includegraphics[width=\linewidth]{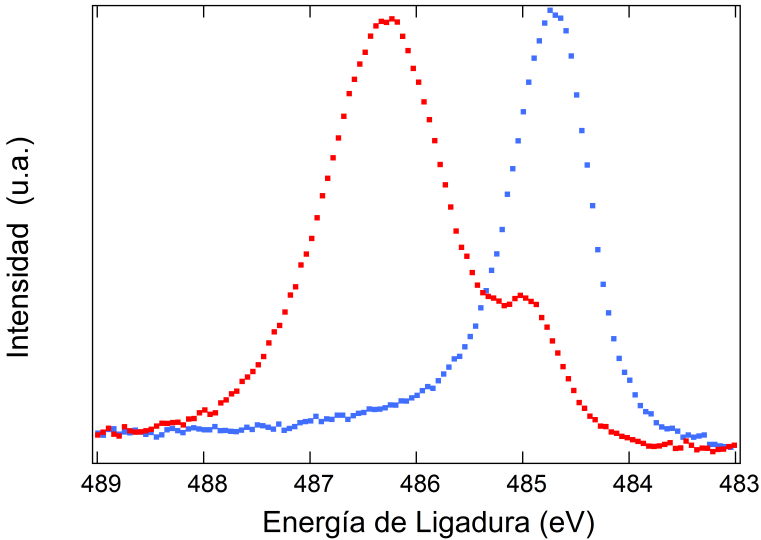}
    \caption{Picos del Sn3d5/2 de la muestra patrón con (1/3+0,4)MC de Sn (en rojo) y de la muestra preparada en el equipo XPS-UPS con 25 minutos netos de evaporación (en azul).}
    \label{fig:conta}
\end{figure}

Como en las sucesivas evaporaciones se notó un ligera presencia de O, el corrimiento del pico de Sn3d5/2 podría atribuirse a la oxidación de algunos átomos de Sn. Sin embargo, en la Figura \ref{fig:conta} se muestra la comparación de los picos del Sn3d5/2 entre el patrón de (1/3+0,4)MC de Sn preparado en el equipo LEED-STM, y entre una muestra preparada en el equipo XPS-UPS con un tiempo de evaporación neto de 25 minutos. La diferencia entre ambos espectros puede explicarse debido a la oxidación de algunos átomos de Sn en la muestra patrón, ya que estuvo en contacto con el O presente en el aire antes de realizarse la medición de XPS. Como el corrimiento por oxidación ($\approx$4 eV) resultó ser un orden de magnitud mayor que el corrimiento medido en la muestra preparada en el equipo XPS-UPS, descartamos que la oxidación sea la culpable de este último. 

\subsection{Análisis de la Relación de Dispersión}

Para determinar la relación de dispersión del estado de superficie de las muestras, se utilizó la técnica del UPS/ARPES. Como en el Laboratorio de Superficies no se han medido sistemáticamente relaciones de dispersión, en primera instancia se procedió a analizar un caso conocido, el Au(111), que además de poseer un estado de superficie fácilmente medible (hecho que permitió ganar experiencia en la técnica del ARPES), la relación de dispersión resulta ser sensible al efecto spin-órbita\cite{LaShell}\cite{Petersen}. Recordemos que en el “estaneno”, el alto valor del número atómico Z=50 del Sn en comparación con el del C, Si y el Ge sería el causante del efecto Hall Cuántico de Spin QSH como se mencionó en la Introducción.
\begin{figure}[ht]
    \centering
    \includegraphics[width=\linewidth]{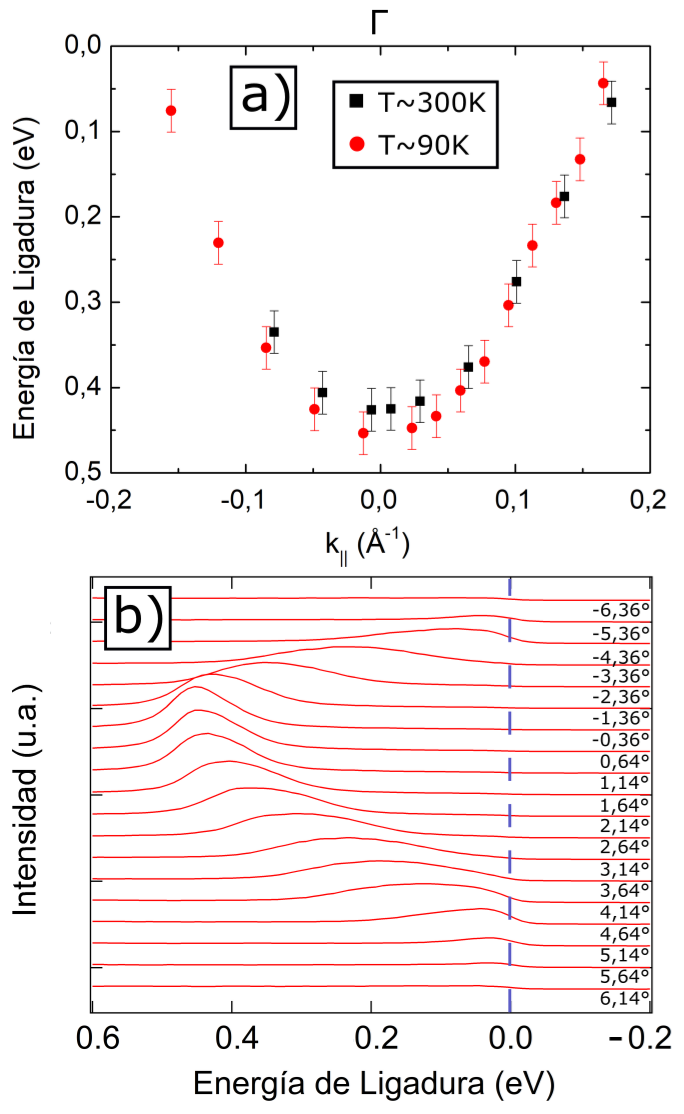}
    \caption{(a) Relación de dispersión del estado de superficie de la muestra de Au(111) a T$\approx$90 K (círculos) y a T$\approx$300K (cuadrados). (b) Espectros de UPS/ARPES del estado de superficie para distintos ángulos de incidencia para T$\approx$90 K. Los ángulos mostrados tienen en cuenta la calibración y además, cada uno de ellos, se corresponde con el espectro inmediatamente superior. Las mediciones se realizaron con una energía de paso de 1 eV, en el modo \textit{Large Area} y con el iris cerrrado.}
    \label{fig:au1}
\end{figure}

\begin{figure}[ht]
    \centering
    \includegraphics[width=\linewidth]{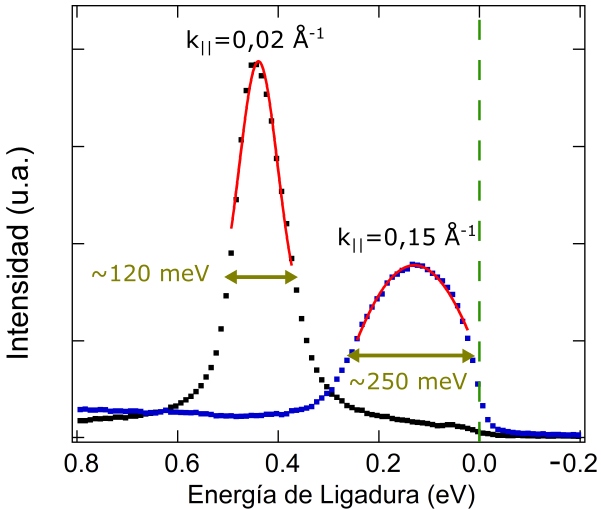}
    \caption{Espectros de ARPES de la muestra de Au(111) para $k_{\parallel}=0,02$ Å$^{-1}$ y para $k_{\parallel}=0,15$ Å$^{-1}$. Puede apreciarse que el ancho del segundo pico resultó ser aproximadamente el doble con respecto al primero.}
    \label{fig:fit_au}
\end{figure}

Se realizaron dos mediciones del estado de superficie de la muestra de Au(111). La primera de ellas fue realizada a temperatura ambiente, mientras que la segunda, fue realizada a aproximadamente 90 K. En la Figura \ref{fig:au1}(a) se observa la relación de dispersión para ambos casos, y en (b) la evolución de los espectros al variar el ángulo de incidencia de la radiación con respecto de la superficie para $T\approx\:90$ K. Ambas curvas coinciden con una relación de dispersión de tipo parabólica, en concordancia con lo medido en \cite{LaShell}. Para elaborar el gráfico de la Figura \ref{fig:au1}(a) primero se determinaron las posiciones en energía de los picos para cada espectro, es decir, para cada ángulo. Las mismas se obtuvieron como resultado de ajustes de tipo Gaussiano en cada uno de los picos (ver Figura \ref{fig:fit_au}). Se estableció como incerteza para la energía el valor obtenido del ancho Gaussiano del ajuste del borde de Fermi a $T\approx\:90$ K. Dicho valor fue del orden de 0,05 eV. Luego, para calibrar el ángulo de incidencia, se realizó el gráfico de la energía de ligadura en función del ángulo de incidencia, se ajustó un función cuadrática y se redefinieron los ángulos teniendo en cuenta la posición en energía del vértice de la parábola obtenida. Posteriormente, en virtud de la Ecuación \ref{eq:ecin} se obtuvo la energía cinética para cada espectro. En el Apéndice se muestra el método utilizado para la determinación de la función trabajo de la muestra de Au(111), $\phi_{Au}$= 5,34 eV. Luego, en virtud de la Ecuación \ref{eq:kpar} se determinó el $k_{\parallel}$ para cada espectro y finalmente se graficó la energía de ligadura en función de dicho parámetro. 

Los picos tienden a ensancharse a medida que el $k_{\parallel}$ crece en módulo, hasta que en la cercanía del borde de Fermi tienden a desaparecer (ver Figuras \ref{fig:au1} y \ref{fig:fit_au}). Por ejemplo, para $k_{\parallel}=0,02$ Å$^{-1}$ el ancho a aproximadamente mitad altura fue del orden de 120 meV, mientras que para $k_{\parallel}=0,15$ Å$^{-1}$ fue del orden de 250 meV. Una posible explicación del fenómeno podría deberse al acoplamiento spín-órbita SOC. En \cite{LaShell} reportan que debido dicho acoplamiento, el pico asociado al estado de superficie se divide y su separación aumenta conforme lo hace $k_{\parallel}$, hasta alcanzar la máxima separación del orden de los 100 meV para $k_{\parallel}\approx0,15$ Å$^{-1}$. Dicho valor es mayor que la resolución que se obtuvo a partir del ajuste del borde de Fermi de $\approx$ 50 meV, y por lo tanto, la resolución en energía sería suficiente para resolver los dos picos asociados al acoplamiento spín-órbita SOC. El hecho de que no se haya podido resolverlos muy probablemente se deba a una resolución angular insuficiente.
\begin{equation}
\Delta E=\frac{dE}{dk_{\parallel}}\Delta k=\frac{dE}{dk_{\parallel}}\frac{dk_{\parallel}}{d\alpha}\Delta \alpha,
\end{equation}
donde $\Delta \alpha$ es la resolución angular del equipo. Para estimarla, puede realizarse un ajuste cuadrático con los datos de la Figura \ref{fig:au1}. Utilizando los parámetros obtenidos del ajuste, se puede calcular la derivada $\frac{dE}{dk_{\parallel}}$. Si por ejemplo, para $k_{\parallel}=0,15$ Å$^{-1}$ se quisiese calcular $\Delta k$ y $\Delta \alpha$ para que $\Delta E\approx$ 250 meV, resulta que $\Delta k$=0,05 Å$^{-1}$ y que $\Delta \alpha$=1,5$^{\circ}$. Por lo tanto, para observar el desdoblamiento del estado de superficie se requeriría una resolución angular menor a 1$^{\circ}$. Las variaciones observadas en los espectros medidos sugieren que la resolución angular del equipo debería ser cercana a 1$^{\circ}$. Entonces, es plausible que el ensanchamiento observado en los espectros al aumentar el ángulo $\theta$ se deba principalmente a la finitud  de la resolución angular. 

\begin{figure}[ht]
    \centering
    \includegraphics[width=0.95\linewidth]{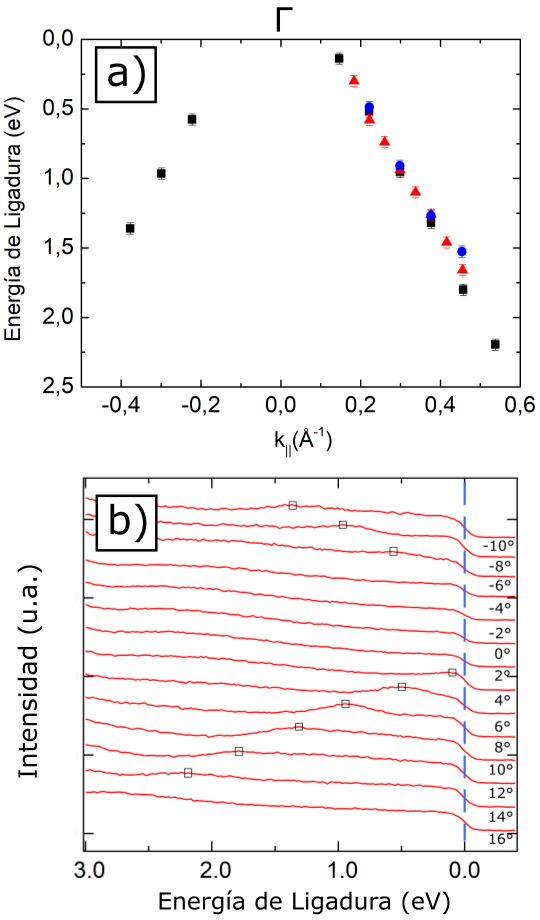}
    \caption{(a) Relación de dispersión del estado de superficie de la muestra de Sn/Ag(111) para 10, 26 y 34 minutos netos de evaporación. (b) Espectros de UPS/ARPES del estado de superficie para distintos ángulos para la muestra con 49 minutos netos de evaporación de Sn. Los ángulos mostrados se corresponden con el espectro inmediatamente superior. Las mediciones se realizaron con una energía de paso de 4 eV, en el modo \textit{Large Area} y con el iris cerrrado.}
    \label{fig:estn_disp}
\end{figure}

Luego, se procedió a analizar el estado de superficie del sistema Sn/Ag(111). En este caso, las mediciones fueron realizadas a temperatura ambiente. En la Figura \ref{fig:estn_disp}(a) se muestra la relación de dispersión del estado de superficie para 10, 26 y 34 minutos netos de evaporación, mientras que en (b) se muestran los espectros medidos para el último caso. Para esta muestra, la función trabajo estimada fue de $\phi_{Sn/Ag}$= 4,57 eV. Las tres mediciones coinciden con una relación de dispersión con forma de $\Lambda$, que resulta ser consistente con la relación de dispersión de la aleación de Ag$_2$Sn\cite{Yuhara}\cite{Jacek}. 
Sin embargo, la dosis de Sn acumalada en la última evaporación es muy próxima a (1/3+0,5)MC de Sn y por lo tanto, debería predominar la banda parabólica reportada en \cite{Yuhara} para la fase del “estaneno”. La explicación de dicha discrepancia podría estar relacionada con la metodología con la cual se preparaba la muestra. 
En vez de calentar la muestra usando el modo radiativo, como se hizo en los experimentos asociados a la Figura \ref{fig:area_tiempo}, en este caso se utilizó el modo  de bombardeo electrónico. Esto podría haber causado un aumento excesivo de la temperatura de la muestra y  en un eventual crecmiento de la aleación Ag$_2$Sn. Esto concuerda con la evolución del pico del estado de superficie a medida que amumentaba el tiempo neto de evaporación de Sn (ver Figura \ref{fig:estn_disp}). En la misma dirección se concluye del análisis del cociente $\epsilon$ que el mismo disminuye considerablemente entre la segunda y tercera evaporación. Estos dos fenómenos podrían ser indicio de que los átomos de Sn se estarían introduciendo en el \textit{bulk}. De ese modo, la progresiva evaporación iría formando una capa de aleación por sobre la anterior, disminuyendo así la cantidad de Sn en la superficie y aumentando la intensidad del pico asociado al estado de superficie. Por lo tanto, recomendamos que para futuras mediciones la muestra de Sn/Ag(111) se calentada más lentamente y utilizando el modo radiativo.

\begin{figure}[ht]
    \centering
    \includegraphics[width=\linewidth]{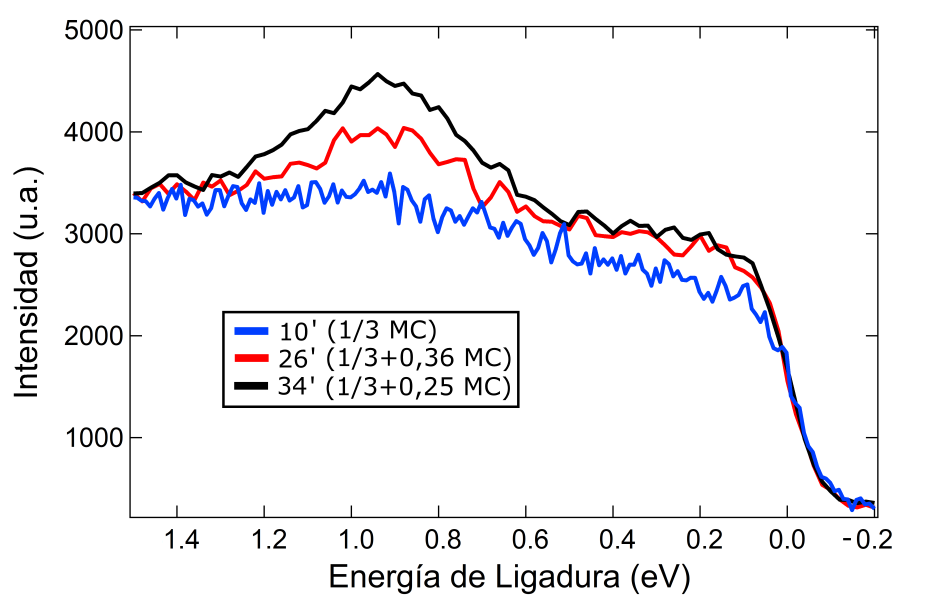}
    \caption{Espectros de UPS/ARPES de la muestra de Sn/Ag(111) con un tiempo de evaporación neto de Sn de 10, 26 y 34 minutos para un ángulo de 8$^{\circ}$.}
    \label{fig:estn_disp}
\end{figure}

\section{\label{sec:Conclusiones}Conclusiones}

Se reprodujeron parcialmente los resultados reportados en \cite{Yuhara} acerca de la formación de “estaneno” sobre un sustrato de Ag(111).

Con LEED se observaron patrones de difracción que se asociaron con una estrucutura de tipo ($\sqrt{3}\times \sqrt{3}$)R30$^{\circ}$ para los átomos de Sn  tanto para 1/3 como para 1/3+0,4 MC de Sn.

Con XPS se observó el corrimiento en energía de los picos asociados al Sn en función de su dosis en la muestra. Para 1/3 MC de Sn los picos pudieron ser ajustados con una sola componente, mientras que para 1/3+0,4 MC de Sn fueron necesarias dos, cuya separación en energía fue del orden de 0,3 eV. Se descartó que el corrimiento haya sido producto de la oxidación de los átomos de Sn.

Con UPS/ARPES se determinaron las relaciones de dispersión tanto de la muestra de Sn/Ag(111) como de la muestra de Au(111). Para el Au(111) se obtuvo una relación de dispersión parabólica y se observó un ensanchamiento en los picos posiblemente debido a la interacción spin-órbita. Mientras que para el Sn/Ag(111) se obtuvo una relación de dispersión con forma de $\Lambda$ para distintas dosis de Sn, la cual resultó muy similar a la relación de dispersión de la aleación de superficie Ag$_2$Sn\cite{Jacek}. Se discutió la influencia del modo de calentamiento en la preparación de la muestra dado que la relación de dispersión no tuvo una forma de tipo parabólica como en \citep{Yuhara}.

\section*{\label{sec:Agrad}Agradecimientos}
El autor quiere agradecerle a los investigadores y al personal de apoyo del Laboratorio de Superfices del Centro Atómico Bariloche por su ayuda y por permitirle realizar la mediciones en sus equipos de trabajo cotidiano. También quiere agradecer personalmente a Hugo por su dedicación y su compromiso a lo largo de todo el trabajo.
\newline

\section* {Apéndice}

\subsection*{Cálculo de la función trabajo de la muestra de Au(111) y de Sn/Ag(111)}

\begin{figure}[H]
    \centering
    \includegraphics[width=\linewidth]{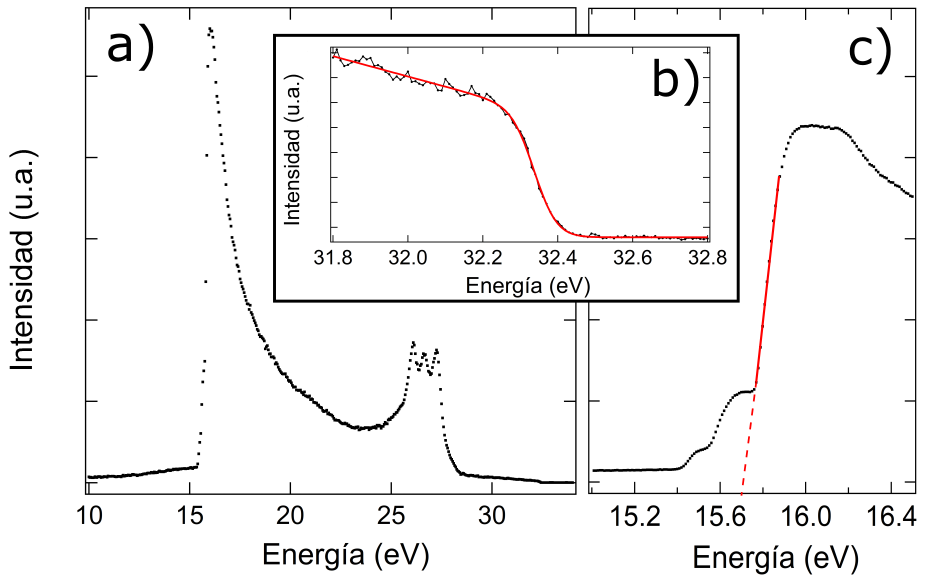}
    \caption{Medición de la Banda de Valencia de la muestra de Sn/Ag(111) con una tensión +12V con respecto de masa. En (a) se observa la Banda de Valencia completa, en (b) el borde de Fermi junto con un ajuste y en (c) el borde asociado a los electrones con la mínima energía cinética posible en la nueva configuración, junto con un ajuste lineal.}
    \label{fig:work_estaneno}
\end{figure}

Para determinar la relación de dispersión de un estado de superficie es necesario conocer el valor de $k_{\parallel}$ para cada ángulo, y por tanto, el valor de la energía cinética. Como la energía cinética se relaciona con la energía de ligadura a partir de una relación que contiene a la función trabajo de la muestra, es necesario determinarla. Para ambas muestras, se aplicó una tensión de +12V entre la muestra y tierra que desplaza a los espectros en energía. En ese estado es posible determinar la energía de Fermi a partir de un ajuste del borde de Fermi y la energía asociada a los electrones con la mínima energía cinética posible en la nueva configuración. Luego, la función trabajo $\phi=h\nu-\Delta$, donde $\Delta$ es la diferencia entre los dos valores de energía antes mencionados. Los valores obtenidos fueron: $\phi_{Au}$= 5,34 eV y $\phi_{Sn/Ag}$= 4,57 eV. En la Figura \ref{fig:work_estaneno} se muestra la medición realizada para el Sn/Ag(111). \\

\begin{thebibliography}{}
\bibitem{Matthew} Matthew J. Allen \textit{et al}. Honeycomb Carbon: A Review of Graphene. \textit{Chem. Rev}. (2010) 110, 132-145. DOI: 10.1021/cr900070d 
\bibitem{Andrew} Andrew J. Mannix\textit{et al}. Synthesis and chemistry of elemental 2D materials. \textit{Nature Reviews - Chemistry}. (2017) Vol. 1 Number 14. DOI; 10.1038/s41570-016-0014.
\bibitem{figStan}Xiaoliang Zhang \textit{et al}. Thermal conductivity of silicene calculated using an optimized Stillinger-Weber potential. \textit{Physical Review} (2014) \textbf{B} 89(5):054310 .DOI:10.1103/PhysRevB.89.054310
\bibitem{aislante}Yimei Fang \textit{et al}. Quantum Spin Hall States in Stanene/Ge(111). \textit{Scientific Reports volume} (2015) \textbf{5}, Article number: 14196. DOI: 10.1038/srep14196 
\bibitem{Zhu}Feng-feng Zhu. Epitaxial growth of two-dimensional stanene. \textit{Nature Materials} (2015) Volume \textbf{14}, pages 1020–1025. DOI:10.1038/nmat4384
\bibitem{Yuhara} Junji Yuhara \textit{et al}. Large area planar stanene epitaxially grown on Ag(111). \textit{2D Mater} (2018) \textbf{5} 025002. DOI:10.1088/2053-1583/aa9ea0
\bibitem{Yeh}J.J. Yeh andI. Lindau. Atomic subshell photoionization cross sections and assymetry parameters $1\leq Z \leq 103$.\textit{ Atomic Data and Nuclear Data Table} (1985) \textbf{32},1-155. DOI:10.1016/0092-640X(85)90016-6
\bibitem{Jacek}Jacek R. Osiecki and R. I. G. Uhrberg. Alloying of Sn in the surface layer of Ag(111). \textit{Physical Review B} (2013) \textbf{87}, 075441.DOI:10.1103/PhysRevB.87.075441 
\bibitem{Doniach}Doniach, S., and Sunjic, M. . Many-electron singularity in X-ray photoemission and X-ray line spectra from metals. \textit{Journal of Physics C: Solid State Physics}, (1970) \textbf{3}(2), 285–291. DOI:10.1088/0022-3719/3/2/010 
\bibitem{LaShell}S. LaShell \textit{et al}. Spin Splitting of an Au(111) Surface State Band Observed with Angle Resolved Photoelectron Spectroscopy.\textit{ Physical Review Letters} (1996) Vol. 77 Number 16. DOI: 10.1103/PhysRevLett.77.3419
\bibitem{Petersen}L. Petersen, P. Hedegard. A simple tight-binding model of spin-orbit splitting of sp-derived surface states. \textit{Surface Science} (2010) 459 49-56. DOI: 10.1016/S0039-6028(00)00441-6

\end{thebibliography}
\end{document}